\documentclass[aps,prl,superscriptaddress,amsmath,amssymb,twocolumn,amsfonts]{revtex4-2}
\usepackage{bm,xcolor,MnSymbol}
\usepackage{graphicx}
\graphicspath{{Pictures/}}
\usepackage{braket,natbib}  
\usepackage[colorlinks=true,linkcolor=magenta,urlcolor=purple,citecolor=magenta,anchorcolor=blue]{hyperref}
\usepackage[normalem]{ulem}
\usepackage{array,multirow}
\usepackage{mathtools}
\usepackage{tabularx,booktabs}

\newcommand{\cn}{\centering}

\begin{document}

\title{Resolving the Thermal Paradox: Many-body localization or fractionalization?}

\author{Saikat Banerjee}
\email{saikat.banerjee@rutgers.edu}
\affiliation{Center for Materials Theory, Rutgers University, Piscataway, New Jersey, 08854, USA}
\author{Piers Coleman}
\email{pcoleman@physics.rutgers.edu}
\affiliation{Center for Materials Theory, Rutgers University, Piscataway, New Jersey, 08854, USA}
\affiliation{Hubbard Theory Consortium, Department of Physics, Royal Holloway, University of London, Egham, Surrey TW20 0EX, UK}

\date{\today}

\begin{abstract}
{Thermal measurements of heat capacity and thermal conductivity in a wide range of insulators and superconductors exhibit a ``thermal paradox": a large linear specific heat reminiscent of neutral  Fermi surfaces in samples that exhibit no corresponding linear temperature coefficient to the thermal conductivity. At first sight, these observations appear to support the formation of a continuum of thermally localized many-body excitations, a form of many-body localization that would be fascinating in its own right. Here, by mapping thermal conductivity measurements onto thermal RC circuits, we argue that the development of extremely long thermal relaxation times, a ``thermal bottleneck," is likely in systems with either many-body localization or neutral Fermi surfaces due to the large ratio between the electron and phonon specific heat capacities. We present a re-evaluation of thermal conductivity measurements in materials exhibiting a thermal paradox that can be used in future experiments to deliberate between these two exciting alternatives.}
\end{abstract}

\maketitle

\textit{Introduction.}\textbf{\---} Several fascinating non-metallic quantum materials (see Table ~\ref{Tab:table_I}) exhibit large linear specific heat capacities $C_V = \gamma T$, \textbf{\---} a hallmark of a neutral Fermi surface~\cite{Baym2008}, yet show no corresponding linear coefficient in the thermal conductivity, $k_0=\kappa/T$, at low temperatures. From the Einstein relation $\kappa = D C_V $ where $D$ is the particulate diffusion constant, $C_V $ is the specific heat capacity~\footnote{The corresponding Einstein relation for the conductivity of a metal is responsible for the Weidemann-Franz ratio between thermal $\kappa$ and electrical conductivity $\sigma$, ${\kappa}/{T} = ( \pi^2 k_B^2/3 e^2) \sigma$} we expect~\cite{Bird2002}
\begin{equation}\label{eq.1}
k_0 = D \gamma.
\end{equation} 
Thus, at first sight, the absence of a thermal conductivity in a system with a linear specific heat suggests the absence of diffusion $D=0$ in a neutral quasiparticle continuum:  a form of many-body localization~\cite{RevModPhys.91.021001}.  In this letter, we identify an important thermal bottleneck which can however obscure any underlying quasiparticle thermal conductivity due to long equilibration times between the phonons and neutral quasiparticles. Our main conclusion is that further measurements, keeping careful track of equilibration times are required to discern between thermal localization and neutral Fermi surfaces. 

Examples of systems exhibiting a thermal paradox are given in Table~\ref{Tab:table_I}). For instance, the Kondo insulator, Samarium Hexaboride (SmB$_6$)  has a linear specific heat in the range $\gamma \sim 5-50$ mJK$^{-2}$mol$^{-1}$~\cite{PhysRevX.4.031012}, considerably larger than the isostructural metal Lanthanum Hexaboride (LaB$_6$), but it exhibits a vanishing residual thermal conductivity at zero magnetic  field~\cite{PhysRevLett.116.246403}. In this particular case, previous experiments have also reported evidence for quantum oscillations in magnetic torque~\cite{Li2014,Tan2015} and specific heat measurements~\cite{Labarre2022} that have been interpreted in terms of excitations above neutral Fermi surfaces with long mean-free paths. Similar paradoxes arise in spin-liquid candidates such as the organic insulator $\beta'$-EtMe$_3$Sb[Pd(dmit)$_2$]$_2$ (dmit-131), which exhibits the absence of a linear thermal conductivity coefficient despite a sizeable linear specific heat~\cite{PhysRevB.105.245133,Yamashita2022}. The case of the heavy fermion superconductor Uranium diTelluride (UTe$_2$) is another interesting example. In this case, samples prepared by chemical vapor deposition display a linear specific heat $\gamma \sim 60$ mJK$^{-2}$mol$^{-1}$ in the superconducting state, one-half the normal state value~\cite{Lewin_2023,Suetsugu2024}. While the linear specific heat disappears in higher-quality flux-grown samples, it is still surprising that samples with half the normal state density of states exhibit no thermal diffusion.  

\renewcommand{\arraystretch}{1.2}
\begin{table*}[t!]
\begin{tabular}{|c|c|c|c|c|c|c|c|c|}
\hline  
System													&		
\multicolumn{2}{c|}{\textbf{Insulator}}					&		
\multicolumn{3}{c|}{\textbf{Quantum spin liquid}}		&	
\multicolumn{1}{c|}{\textbf{Superconductor}}	\\
\hline \hline
\cn		                    &		
SmB$_{6}$					&		
YbB$_{12}$					&		
dmit-131					&
NaYbS$_2$, NaYbSe$_2$		&	
1T-TaS$_2$					&	
UTe$_{2}$					\\
\hline
Ref		&	
\cite{PhysRevX.4.031012,PhysRevLett.116.246403,PhysRevB.99.045138}	&	
\cite{Sato2019}														&	
\cite{Yamashita2011,Yamashita2022,PhysRevX.9.041051}				&
\cite{PhysRevB.110.224414,PhysRevB.100.224417,PhysRevX.11.021044}	&
\cite{PhysRevB.96.081111,Kratochvilova2017}				            &	
\cite{Imajo2019,Lewin_2023,Suetsugu2024,Ishihara2023}							\\
\hline
$\gamma$ (mJK$^{-2}$ mol$^{-1}$)	&
$\sim 25$							&
$\sim 3.8$							&
$\sim 20$							&
$\sim 1.1$							&
$\sim 1.84$							&
$\sim 60$							\\
\hline
$\beta$	(mJK$^{-4}$mol$^{-1}$)	&
$\sim 0.45$  					&
$\sim 0.026$ 					&
$\sim 24$	 					&		
$\sim 0.1$ $^\ast$		        &
$\sim 0.31$						&
$\sim 2.81$    					\\
\hline
$C_f/C_p=\gamma/\beta T_0^2$ 		& 	
$\sim 350$							&
$\sim 584$							&
$\sim 40$							&
$\sim 1000$							&		
$\sim 590$							&
$\sim 133$							\\
\hline
Base temperature $T_0$ (K)			& 	
$\sim 0.4$  						&
$\sim 0.5$	 						&
$\sim 0.15$							&
$\sim 0.1$							&		
$\sim 0.1$							&
$\sim 0.4$ 							\\
\hline
$\kappa_0/T$ (mWK$^{-2}$cm$^{-1}$)	& 	
$\sim 0$  							&
$\sim 0.01$	 						&
$\sim 0$							&
$\sim 0$							&		
$\sim 0$							&
$\sim 0$ 							\\
\hline 
\end{tabular}
\caption{Measured values of the various coefficients in the specific heat capacity $C/T = \gamma + \beta T^2$, ($\gamma$ is typically attributed to spin/fermions, and $\beta$ to phonons) and residual thermal conductivity $\kappa_0/T$ at low temperatures. $C_f/C_p=\gamma/\beta T_0^2$  gives a ratio of the spin/fermion to phonon specific heat at the base measurement temperature $T_0$. ($^\ast$The phonon contribution is measured from the isostructural compound NaYbO$_2$~\cite{Bordelon2019}). 
}
\label{Tab:table_I}
\end{table*}

Yamashita et al.~\cite{Yamashita2022} \--- have recently noted that the measured low-temperature thermal conductivity depends on the cooling rate used in the experiment. This suggests that thermal equilibrium in the temperature gradient is not established without a very slow measurement protocol.  This conclusion is corroborated by more recent measurements on dmit-131~\cite{PhysRevX.9.041051}, which hint at a subtle spin-phonon coupling responsible for the slow transmission of heat into the underlying spin fluid.  
 

To investigate thermal relaxation rates from various possible sources, we map the problem onto a thermal circuit diagram~\cite{Vollmer_2009}. The sample is modeled as a one-dimensional object with endpoints at two distinct temperatures. The steady-state heat thermal current between its two ends, $i$ and $j$,  is described by the thermal Ohm's law
\begin{equation}\label{eq.2}
I_T(i\rightarrow j)  = \left(\frac{\kappa A}{l}\right)(T_i-T_j)
\end{equation}
where $I_T$ is the heat current flowing from $i$ to $j$, $T_i-T_j$ is the difference in temperatures, $\kappa$ is the thermal conductivity, $A$ and $l$ are the cross-sectional area and length of the sample, respectively and  $\kappa$ is the total thermal conductivity, \textit{i.e.}, an aggregate of contributions from various channels such as phonons, electrons and other quasiparticles  $\kappa = \kappa_{\rm{p}} + \kappa_e + \ldots$. The conservation of heat in a thermal circuit allows us to exploit the analogy between heat and charge $Q_T\leftrightarrow Q$, thermal and electrical current $I_T\leftrightarrow I_e$. We can identify temperature with voltage  ($T_i\leftrightarrow V_i$) and $R_i = l / (\kappa_i A) $ as the thermal resistance of a given thermal conduction channel. In a similar fashion, we can identify a heat bath with a capacitor in which the heating rate is equal to the incoming thermal current $dQ_T/dt = I_T$: by comparing $dQ_T/dT = C_T$ with $dQ/dV = C_e$, we may identify the specific heat $C_T= C_e$ as a thermal capacitance. These connections allow us to represent heat conduction circuits as RC electrical circuits~\cite{Kakac1985,Mills1992,Chen2015}, allowing us to analyze various thermal relaxation timescales in terms of the corresponding time constants $\tau \sim C_T R_T$, as summarized in Table \ref{Tab:table_II}.
Curiously, thermal RC circuits have not been widely used in a condensed matter framework, but they have been extensively used to model heat transfer in urban buildings and climate sciences~\cite{Bueno2012,Silva2022}.

\renewcommand{\arraystretch}{1.5}
\begin{table}[b!]
\begin{tabular}{|c|c|c|}
\hline  
			&	\textbf{Electrical}						&		\textbf{Thermal}		\\
\hline
Quantity	&	$Q=$ Charge								&		$Q_T=$ Heat				\\
\hline
Potential	&	$V$ (voltage)							&		$T$ (temperature)		\\
\hline 
Ohm's Law	&	$V(\omega ) = I(\omega) Z(\omega)$ 		&		$T(s) = I_T(s) Z(s)$	\\
\hline
$Z_R$		&	$R_e $									&		$R_T = \frac{l}{A \kappa}$	\\
\hline
$Z_C$		&	$\frac{1}{-i \omega C}$					&		$\frac{1}{- s C}$		\\
\hline
\end{tabular}
\caption{Mapping between electrical and thermal circuits.}\label{Tab:table_II}
\end{table}

\begin{figure}[b!]
\centering
\includegraphics[width=0.85\linewidth]{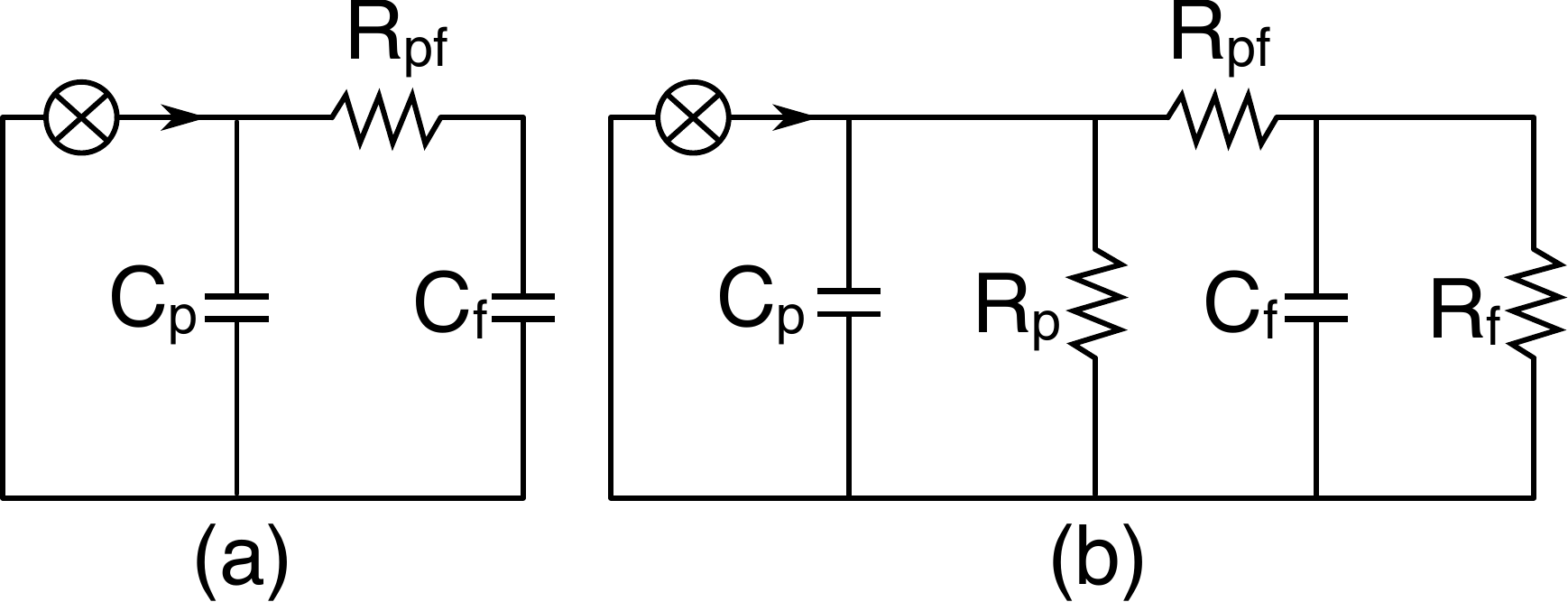}
\caption{Thermal circuit diagrams describing  (a) specific heat and (b) thermal conductivity measurements.}\label{fig:Fig1}
\end{figure}

\textit{Model.}\textbf{\---} We consider the effect of weak thermal coupling between the phonon and spin/fermion degrees of freedom on the equilibration times in a specific heat and a thermal conductivity measurement, introducing two circuit models, depicted in Fig.~\ref{fig:Fig1}(a,b). Fig.~\ref{fig:Fig1}(a) illustrates the circuit for a specific heat measurement. In a specific heat measurement, a pulse of heat is sent into the thermally isolated system.  The heat initially flows into the phonon modes of the material, represented by capacitance $C_{\rm p}$. As the lattice temperature rises, heat is transferred into the electronic and magnetic modes of the material via the thermal resistance $R_{\rm pf}$, flowing into the spin/fermion degrees of freedom, represented by capacitance $C_{\rm f}$. Fig.~\ref{fig:Fig1}(b) shows the corresponding circuit for thermal conductivity measurement. In this case, a constant heat flux is applied to one surface, measuring the steady-state temperature gradient once equilibrium is established. The heat initially accumulates in the phonon modes, represented by $C_{\rm{p}}$, and is then transmitted into the electronic and magnetic modes via the thermal resistance $R_{\rm{p}}$. The spin/fermion degrees of freedom conduct heat via a thermal resistance $R_{\rm{f}}$ which discharges the heat in the spin-fermion system carried by thermal capacitance $C_{\rm{f}}$. If many-body localized spin/fermion degrees of freedom are present,  $R_{\rm{f}}$ is essentially infinite. Although these models provide a simplified representation of the measurement protocols, they capture the key phenomenological features of thermal relaxation. 

In a canonical insulator, the low-temperature phonon specific heat is a cubic function of temperature; we are particularly  interested in the case where the electronic/spin degrees of freedom exhibit a linear specific heat so that 
\begin{equation}\label{eq.3}
C_{\rm{f}} = \gamma T, \quad
C_{\rm{p}} = \beta T^3.
\end{equation}
Experimental estimates of $\beta,\gamma$ for various materials are provided in Table~\ref{Tab:table_I}. 

\textit{Impedance and temperature profile.}\textbf{\---} Table \ref{Tab:table_I} summarizes the analogy between electrical and thermal circuits.  Since thermal equilibration is most conveniently described in terms of Laplace transforms, we write the temperature as transformed as follows 
$$T(s) = \int_0^{\infty} T(t) e^{-st} dt.$$
In the Laplace transform  $ s$ plays the role of an imaginary frequency, and in the generalized thermal Ohm's law  $T(s) =I_T(s) Z(s)$ we must then replace $i\omega \rightarrow s$ in each of the circuit elements, so that a heat bath, as a thermal capacitor is associated with an impedance $Z(s) = \frac{1}{-s C}$.

We now compute the impedances for each of the circuits depicted in Fig.~\ref{fig:Fig1}(a,b). A key advantage of this analysis is that the location of the poles $s_l = i \Gamma_l$ determines the thermal equilibration rates $\Gamma_l= 1/\tau_l$ of the circuit.  For specific heat measurements, the thermal impedance (see Supplementary Material (SM)~\cite{supp} for details) involves a single pole with the time constant as 
\begin{equation}\label{eq.4}
\frac{1}{\tau_0} = \frac{1}{R_{\rm{pf}}} \left( \frac{1}{C_{\rm{p}}} + \frac{1}{C_{\rm{f}}}\right).
\end{equation}
In the limit of large spin/fermion capacitance ($\gamma \gg \beta T_0^2$, where $T_0$ is the characteristic temperature scale for the measurement, see Table~\ref{Tab:table_I}), the time-constant becomes $\tau_0 = R_{\rm{pf}} C_{\rm{p}}$, indicating that the large thermal resistances of the spin fluid and phonon do not affect the time required for heat capacity measurement. 

However, in a thermal conductivity measurement, modeled by the circuit diagram in Fig.~\ref{fig:Fig1}(b), the thermal resistances for both the spin/fermion and phonon play a role in the equilibration, leading to  fast and a slow time constants
\begin{equation}\label{eq.5}
\tau_{\rm{F}} =  \frac{C_{\rm{p}} R_{\rm{pf}} R_{\rm{p}}}{R_{\rm{pf}} + R_{\rm{p}}}, \quad 
\tau_{\rm{S}} = \frac{C_{\rm{f}} R_{\rm{f}} (R_{\rm{p}} + R_{\rm{pf}})}{R_{\rm{pf}} + R_{\rm{p}} + R_{\rm{f}}}.
\end{equation}
It is the large ratio between the specific heat of the electron and phonon fluids that sets the ratio between $\tau_{\rm S}/\tau_{\rm F} \sim C_{\rm f} /C_{\rm p}\gg 1$. 
Whereas the fast time-constant $\tau_{\rm{F}}$ is even shorter than the equilibration time of a specific heat measurement,  the slow time-constant $\tau_{\rm{S}}$ is much larger than both $\tau_0$ and $\tau_{\rm{F}}$. We can understand these very different time scales by noting that at the short times governing $\tau_{\rm S}$, the thermal capacitance $C_{\rm{f}}$  is shorted out, while at the long-times governing $\tau_{S}$,  $C_{\rm{f}}$ is in series with the parallel resistors $R_{\rm{pf}} + R_{\rm{p}}$, and $R_{\rm{f}}$. In the limit of large spin/fermion thermal resistance ($R_{\rm{f}} \to \infty$), the fast time-constant $\tau_{\rm{F}}$ remains the same, while the slow time-constant $\tau_{\rm{S}}$ becomes much longer. Note that within our model $R_{\rm{f}} \to \infty$ would serve as a potential signature of many-body localization.

\begin{figure}[t!]
\centering
\includegraphics[width=0.8\linewidth]{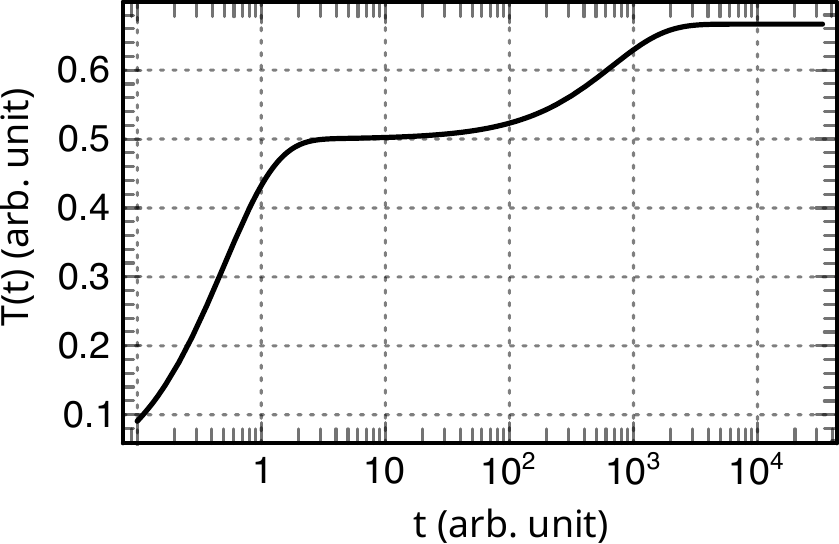}
\caption{An illustration of the time-dependent temperature profile in case of thermal conductivity measurement, obtained by utilizing Eq.~\eqref{eq.6}. For quantitative purposes, we adopt $C_{\rm{p}} = 1$, $C_{\rm{f}} = 1000$, and $R_{\rm{p}}= R_{\rm{pf}} = R_{\rm{f}} = 1$ in arbitrary unit. The ratio $C_{\rm{f}}: C_{\rm{p}}$ is chosen as a large value motivated by the experimental estimates for specific heat capacity ($\gamma$, and $\beta$) in various materials, see Table.~\ref{Tab:table_I}.} \label{fig:Fig2}
\end{figure}

The temperature profile as a function of time in each of these circuits is obtained by inverting the Laplace transform $T(s)= I(s)Z(s)$ back into the time domain, which yields 
\begin{equation}\label{eq.6}
T(t) = I_0 \biggl[A_1 (1-e^{-t/\tau_{\rm{F}}}) + A_2(1-e^{-t/\tau_{\rm{S}}}) \biggr],
\end{equation}
where $A_1, A_2$ are constants depending on the thermal resistances and capacitances defined in Fig.~\ref{fig:Fig1}(b), and $I_0$ is the long-time equilibration thermal current. A detailed derivation can be found in SM~\cite{supp}. The corresponding temperature evolution relevant for the thermal conductivity measurement is shown in Fig.~\ref{fig:Fig2} for a very large spin fluid capacitance ($C_{\rm{f}} \gg C_{\rm{p}}$), and assuming almost identical thermal resistances ($R_{\rm{p}} \sim R_{\rm{pf}} \sim R_{\rm{f}}$). It is clear from Fig.~\ref{fig:Fig2} that the large capacitance $C_{\rm{f}}$ only affects the time constant for measuring thermal conductivity, not the time constant for measuring the specific heat.

In a steady-state measurement, thermal capacitances can be neglected under the assumption that they are fully charged. Under this condition, we can estimate the secondary temperature increment predicted in Fig.~\ref{fig:Fig2} using the circuit representation in Fig.~\ref{fig:Fig1}(a). The temperature difference is   determined from the thermal resistances as
\begin{equation}\label{eq.7}
\frac{1}{\Delta T_{\rm{F}}} - \frac{1}{\Delta T_{\rm{I}}} 
= 
\frac{1}{I_0}
\left(
\frac{1}{R_{\rm{pf}}} - \frac{1}{R_{\rm{pf}}+R_{\rm{f}}}
\right),
\end{equation}
where $\Delta T_{\rm{I}}$ ($\Delta T_{\rm{F}}$) is the initial (final) temperature increment as visualized in Fig.~\ref{fig:Fig2}. This simple result shows the necessity of long waiting times to determine whether the underlying system contains neutral quasiparticles or many-body localized excitations.

\textit{Discussion and conclusion.}\textbf{\---} The paradox of coexisting linear specific heat with negligible residual thermal conductivity in various quantum materials signals a need for a deeper understanding of heat transport mechanisms in systems with neutral fermionic excitations. While the lack of thermal diffusion may initially suggest evidence of many-body localization, our analysis suggests that slow thermal equilibration \--- primarily due to weak phonon-quasiparticle coupling \--- could serve as an alternative explanation. 

By modeling thermal transport in these systems using an electrical circuit analogy, we have identified key relaxation timescales that may obscure the intrinsic heat conduction properties of the spin and fermion degrees of freedom. Our findings indicate that it is crucial to reconsider current measurement protocols, particularly in relation to equilibration times, to distinguish between true thermal localization and hidden neutral Fermi surfaces. Further experimental investigations, including time-resolved thermal transport measurements, are necessary to fully understand the nature of these low-temperature excitations across these class of materials.

\textit{Acknowledgments.}\textbf{\---} This work is supported by the Office of Basic Energy Sciences, Material Sciences and Engineering Division, U.S. Department of Energy (DOE) under Contracts  No. DE-FG02-99ER45790.

\bibliography{References}

\begin{thebibliography}{32}%
\makeatletter
\providecommand \@ifxundefined [1]{%
 \@ifx{#1\undefined}
}%
\providecommand \@ifnum [1]{%
 \ifnum #1\expandafter \@firstoftwo
 \else \expandafter \@secondoftwo
 \fi
}%
\providecommand \@ifx [1]{%
 \ifx #1\expandafter \@firstoftwo
 \else \expandafter \@secondoftwo
 \fi
}%
\providecommand \natexlab [1]{#1}%
\providecommand \enquote  [1]{``#1''}%
\providecommand \bibnamefont  [1]{#1}%
\providecommand \bibfnamefont [1]{#1}%
\providecommand \citenamefont [1]{#1}%
\providecommand \href@noop [0]{\@secondoftwo}%
\providecommand \href [0]{\begingroup \@sanitize@url \@href}%
\providecommand \@href[1]{\@@startlink{#1}\@@href}%
\providecommand \@@href[1]{\endgroup#1\@@endlink}%
\providecommand \@sanitize@url [0]{\catcode `\\12\catcode `\$12\catcode
  `\&12\catcode `\#12\catcode `\^12\catcode `\_12\catcode `\%12\relax}%
\providecommand \@@startlink[1]{}%
\providecommand \@@endlink[0]{}%
\providecommand \url  [0]{\begingroup\@sanitize@url \@url }%
\providecommand \@url [1]{\endgroup\@href {#1}{\urlprefix }}%
\providecommand \urlprefix  [0]{URL }%
\providecommand \Eprint [0]{\href }%
\providecommand \doibase [0]{https://doi.org/}%
\providecommand \selectlanguage [0]{\@gobble}%
\providecommand \bibinfo  [0]{\@secondoftwo}%
\providecommand \bibfield  [0]{\@secondoftwo}%
\providecommand \translation [1]{[#1]}%
\providecommand \BibitemOpen [0]{}%
\providecommand \bibitemStop [0]{}%
\providecommand \bibitemNoStop [0]{.\EOS\space}%
\providecommand \EOS [0]{\spacefactor3000\relax}%
\providecommand \BibitemShut  [1]{\csname bibitem#1\endcsname}%
\let\auto@bib@innerbib\@empty
\bibitem [{\citenamefont {Baym}\ and\ \citenamefont
  {Pethick}(2008)}]{Baym2008}%
  \BibitemOpen
  \bibfield  {author} {\bibinfo {author} {\bibfnamefont {G.}~\bibnamefont
  {Baym}}\ and\ \bibinfo {author} {\bibfnamefont {C.}~\bibnamefont {Pethick}},\
  }\href {https://books.google.com/books?id=xmiV4YSEjE4C} {\emph {\bibinfo
  {title} {Landau Fermi-Liquid Theory: Concepts and Applications}}}\ (\bibinfo
  {publisher} {Wiley},\ \bibinfo {year} {2008})\BibitemShut {NoStop}%
\bibitem [{Note1()}]{Note1}%
  \BibitemOpen
  \bibinfo {note} {The corresponding Einstein relation for the conductivity of
  a metal is responsible for the Weidemann-Franz ratio between thermal $\kappa
  $ and electrical conductivity $\sigma $, ${\kappa }/{T} = ( \pi ^2 k_B^2/3
  e^2) \sigma $}\BibitemShut {NoStop}%
\bibitem [{\citenamefont {Bird}\ \emph {et~al.}(2002)\citenamefont {Bird},
  \citenamefont {Stewart},\ and\ \citenamefont {Lightfoot}}]{Bird2002}%
  \BibitemOpen
  \bibfield  {author} {\bibinfo {author} {\bibfnamefont {R.}~\bibnamefont
  {Bird}}, \bibinfo {author} {\bibfnamefont {W.}~\bibnamefont {Stewart}},\ and\
  \bibinfo {author} {\bibfnamefont {E.}~\bibnamefont {Lightfoot}},\ }\href
  {https://books.google.com/books?id=wYnRQwAACAAJ} {\emph {\bibinfo {title}
  {Transport Phenomena}}}\ (\bibinfo  {publisher} {J. Wiley},\ \bibinfo {year}
  {2002})\BibitemShut {NoStop}%
\bibitem [{\citenamefont {Abanin}\ \emph {et~al.}(2019)\citenamefont {Abanin},
  \citenamefont {Altman}, \citenamefont {Bloch},\ and\ \citenamefont
  {Serbyn}}]{RevModPhys.91.021001}%
  \BibitemOpen
  \bibfield  {author} {\bibinfo {author} {\bibfnamefont {D.~A.}\ \bibnamefont
  {Abanin}}, \bibinfo {author} {\bibfnamefont {E.}~\bibnamefont {Altman}},
  \bibinfo {author} {\bibfnamefont {I.}~\bibnamefont {Bloch}},\ and\ \bibinfo
  {author} {\bibfnamefont {M.}~\bibnamefont {Serbyn}},\ }\bibfield  {title}
  {\bibinfo {title} {Colloquium: Many-body localization, thermalization, and
  entanglement},\ }\href {https://doi.org/10.1103/RevModPhys.91.021001}
  {\bibfield  {journal} {\bibinfo  {journal} {Rev. Mod. Phys.}\ }\textbf
  {\bibinfo {volume} {91}},\ \bibinfo {pages} {021001} (\bibinfo {year}
  {2019})}\BibitemShut {NoStop}%
\bibitem [{\citenamefont {Phelan}\ \emph {et~al.}(2014)\citenamefont {Phelan},
  \citenamefont {Koohpayeh}, \citenamefont {Cottingham}, \citenamefont
  {Freeland}, \citenamefont {Leiner}, \citenamefont {Broholm},\ and\
  \citenamefont {McQueen}}]{PhysRevX.4.031012}%
  \BibitemOpen
  \bibfield  {author} {\bibinfo {author} {\bibfnamefont {W.~A.}\ \bibnamefont
  {Phelan}}, \bibinfo {author} {\bibfnamefont {S.~M.}\ \bibnamefont
  {Koohpayeh}}, \bibinfo {author} {\bibfnamefont {P.}~\bibnamefont
  {Cottingham}}, \bibinfo {author} {\bibfnamefont {J.~W.}\ \bibnamefont
  {Freeland}}, \bibinfo {author} {\bibfnamefont {J.~C.}\ \bibnamefont
  {Leiner}}, \bibinfo {author} {\bibfnamefont {C.~L.}\ \bibnamefont
  {Broholm}},\ and\ \bibinfo {author} {\bibfnamefont {T.~M.}\ \bibnamefont
  {McQueen}},\ }\bibfield  {title} {\bibinfo {title} {{Correlation between Bulk
  Thermodynamic Measurements and the Low-Temperature-Resistance Plateau in
  SmB$_6$}},\ }\href {https://doi.org/10.1103/PhysRevX.4.031012} {\bibfield
  {journal} {\bibinfo  {journal} {Phys. Rev. X}\ }\textbf {\bibinfo {volume}
  {4}},\ \bibinfo {pages} {031012} (\bibinfo {year} {2014})}\BibitemShut
  {NoStop}%
\bibitem [{\citenamefont {Xu}\ \emph {et~al.}(2016)\citenamefont {Xu},
  \citenamefont {Cui}, \citenamefont {Dong}, \citenamefont {Zhao},
  \citenamefont {Wu}, \citenamefont {Chen}, \citenamefont {Sun}, \citenamefont
  {Yao},\ and\ \citenamefont {Li}}]{PhysRevLett.116.246403}%
  \BibitemOpen
  \bibfield  {author} {\bibinfo {author} {\bibfnamefont {Y.}~\bibnamefont
  {Xu}}, \bibinfo {author} {\bibfnamefont {S.}~\bibnamefont {Cui}}, \bibinfo
  {author} {\bibfnamefont {J.~K.}\ \bibnamefont {Dong}}, \bibinfo {author}
  {\bibfnamefont {D.}~\bibnamefont {Zhao}}, \bibinfo {author} {\bibfnamefont
  {T.}~\bibnamefont {Wu}}, \bibinfo {author} {\bibfnamefont {X.~H.}\
  \bibnamefont {Chen}}, \bibinfo {author} {\bibfnamefont {K.}~\bibnamefont
  {Sun}}, \bibinfo {author} {\bibfnamefont {H.}~\bibnamefont {Yao}},\ and\
  \bibinfo {author} {\bibfnamefont {S.~Y.}\ \bibnamefont {Li}},\ }\bibfield
  {title} {\bibinfo {title} {{Bulk Fermi Surface of Charge-Neutral Excitations
  in SmB$_6$ or Not: A Heat-Transport Study}},\ }\href
  {https://doi.org/10.1103/PhysRevLett.116.246403} {\bibfield  {journal}
  {\bibinfo  {journal} {Phys. Rev. Lett.}\ }\textbf {\bibinfo {volume} {116}},\
  \bibinfo {pages} {246403} (\bibinfo {year} {2016})}\BibitemShut {NoStop}%
\bibitem [{\citenamefont {Li}\ \emph {et~al.}(2014)\citenamefont {Li},
  \citenamefont {Xiang}, \citenamefont {Yu}, \citenamefont {Asaba},
  \citenamefont {Lawson}, \citenamefont {Cai}, \citenamefont {Tinsman},
  \citenamefont {Berkley}, \citenamefont {Wolgast}, \citenamefont {Eo},
  \citenamefont {Kim}, \citenamefont {Kurdak}, \citenamefont {Allen},
  \citenamefont {Sun}, \citenamefont {Chen}, \citenamefont {Wang},
  \citenamefont {Fisk},\ and\ \citenamefont {Li}}]{Li2014}%
  \BibitemOpen
  \bibfield  {author} {\bibinfo {author} {\bibfnamefont {G.}~\bibnamefont
  {Li}}, \bibinfo {author} {\bibfnamefont {Z.}~\bibnamefont {Xiang}}, \bibinfo
  {author} {\bibfnamefont {F.}~\bibnamefont {Yu}}, \bibinfo {author}
  {\bibfnamefont {T.}~\bibnamefont {Asaba}}, \bibinfo {author} {\bibfnamefont
  {B.}~\bibnamefont {Lawson}}, \bibinfo {author} {\bibfnamefont
  {P.}~\bibnamefont {Cai}}, \bibinfo {author} {\bibfnamefont {C.}~\bibnamefont
  {Tinsman}}, \bibinfo {author} {\bibfnamefont {A.}~\bibnamefont {Berkley}},
  \bibinfo {author} {\bibfnamefont {S.}~\bibnamefont {Wolgast}}, \bibinfo
  {author} {\bibfnamefont {Y.~S.}\ \bibnamefont {Eo}}, \bibinfo {author}
  {\bibfnamefont {D.-J.}\ \bibnamefont {Kim}}, \bibinfo {author} {\bibfnamefont
  {C.}~\bibnamefont {Kurdak}}, \bibinfo {author} {\bibfnamefont {J.~W.}\
  \bibnamefont {Allen}}, \bibinfo {author} {\bibfnamefont {K.}~\bibnamefont
  {Sun}}, \bibinfo {author} {\bibfnamefont {X.~H.}\ \bibnamefont {Chen}},
  \bibinfo {author} {\bibfnamefont {Y.~Y.}\ \bibnamefont {Wang}}, \bibinfo
  {author} {\bibfnamefont {Z.}~\bibnamefont {Fisk}},\ and\ \bibinfo {author}
  {\bibfnamefont {L.}~\bibnamefont {Li}},\ }\bibfield  {title} {\bibinfo
  {title} {{Two-dimensional Fermi surfaces in Kondo insulator SmB$_6$}},\
  }\href {https://doi.org/10.1126/science.1250366} {\bibfield  {journal}
  {\bibinfo  {journal} {Science}\ }\textbf {\bibinfo {volume} {346}},\ \bibinfo
  {pages} {1208} (\bibinfo {year} {2014})}\BibitemShut {NoStop}%
\bibitem [{\citenamefont {Tan}\ \emph {et~al.}(2015)\citenamefont {Tan},
  \citenamefont {Hsu}, \citenamefont {Zeng}, \citenamefont {Hatnean},
  \citenamefont {Harrison}, \citenamefont {Zhu}, \citenamefont {Hartstein},
  \citenamefont {Kiourlappou}, \citenamefont {Srivastava}, \citenamefont
  {Johannes}, \citenamefont {Murphy}, \citenamefont {Park}, \citenamefont
  {Balicas}, \citenamefont {Lonzarich}, \citenamefont {Balakrishnan},\ and\
  \citenamefont {Sebastian}}]{Tan2015}%
  \BibitemOpen
  \bibfield  {author} {\bibinfo {author} {\bibfnamefont {B.~S.}\ \bibnamefont
  {Tan}}, \bibinfo {author} {\bibfnamefont {Y.-T.}\ \bibnamefont {Hsu}},
  \bibinfo {author} {\bibfnamefont {B.}~\bibnamefont {Zeng}}, \bibinfo {author}
  {\bibfnamefont {M.~C.}\ \bibnamefont {Hatnean}}, \bibinfo {author}
  {\bibfnamefont {N.}~\bibnamefont {Harrison}}, \bibinfo {author}
  {\bibfnamefont {Z.}~\bibnamefont {Zhu}}, \bibinfo {author} {\bibfnamefont
  {M.}~\bibnamefont {Hartstein}}, \bibinfo {author} {\bibfnamefont
  {M.}~\bibnamefont {Kiourlappou}}, \bibinfo {author} {\bibfnamefont
  {A.}~\bibnamefont {Srivastava}}, \bibinfo {author} {\bibfnamefont {M.~D.}\
  \bibnamefont {Johannes}}, \bibinfo {author} {\bibfnamefont {T.~P.}\
  \bibnamefont {Murphy}}, \bibinfo {author} {\bibfnamefont {J.-H.}\
  \bibnamefont {Park}}, \bibinfo {author} {\bibfnamefont {L.}~\bibnamefont
  {Balicas}}, \bibinfo {author} {\bibfnamefont {G.~G.}\ \bibnamefont
  {Lonzarich}}, \bibinfo {author} {\bibfnamefont {G.}~\bibnamefont
  {Balakrishnan}},\ and\ \bibinfo {author} {\bibfnamefont {S.~E.}\ \bibnamefont
  {Sebastian}},\ }\bibfield  {title} {\bibinfo {title} {{Unconventional Fermi
  surface in an insulating state}},\ }\href
  {https://doi.org/10.1126/science.aaa7974} {\bibfield  {journal} {\bibinfo
  {journal} {Science}\ }\textbf {\bibinfo {volume} {349}},\ \bibinfo {pages}
  {287} (\bibinfo {year} {2015})}\BibitemShut {NoStop}%
\bibitem [{\citenamefont {LaBarre}\ \emph {et~al.}(2021)\citenamefont
  {LaBarre}, \citenamefont {Rydh}, \citenamefont {Palmer-Fortune},
  \citenamefont {Frothingham}, \citenamefont {Hannahs}, \citenamefont
  {Ramirez},\ and\ \citenamefont {Fortune}}]{Labarre2022}%
  \BibitemOpen
  \bibfield  {author} {\bibinfo {author} {\bibfnamefont {P.~G.}\ \bibnamefont
  {LaBarre}}, \bibinfo {author} {\bibfnamefont {A.}~\bibnamefont {Rydh}},
  \bibinfo {author} {\bibfnamefont {J.}~\bibnamefont {Palmer-Fortune}},
  \bibinfo {author} {\bibfnamefont {J.~A.}\ \bibnamefont {Frothingham}},
  \bibinfo {author} {\bibfnamefont {S.~T.}\ \bibnamefont {Hannahs}}, \bibinfo
  {author} {\bibfnamefont {A.~P.}\ \bibnamefont {Ramirez}},\ and\ \bibinfo
  {author} {\bibfnamefont {N.}~\bibnamefont {Fortune}},\ }\bibfield  {title}
  {\bibinfo {title} {{Observation of Quantum Oscillations in The Low
  Temperature Specific Heat of SmB$_6$}},\ }\href {arXiv:2111.03758} {\bibfield
   {journal} {\bibinfo  {journal} {preprint}\ ,\ \bibinfo {pages}
  {arXiv:2111.03758}} (\bibinfo {year} {2021})}\BibitemShut {NoStop}%
\bibitem [{\citenamefont {Nomoto}\ \emph {et~al.}(2022)\citenamefont {Nomoto},
  \citenamefont {Yamashita}, \citenamefont {Akutsu}, \citenamefont {Nakazawa},\
  and\ \citenamefont {Kato}}]{PhysRevB.105.245133}%
  \BibitemOpen
  \bibfield  {author} {\bibinfo {author} {\bibfnamefont {T.}~\bibnamefont
  {Nomoto}}, \bibinfo {author} {\bibfnamefont {S.}~\bibnamefont {Yamashita}},
  \bibinfo {author} {\bibfnamefont {H.}~\bibnamefont {Akutsu}}, \bibinfo
  {author} {\bibfnamefont {Y.}~\bibnamefont {Nakazawa}},\ and\ \bibinfo
  {author} {\bibfnamefont {R.}~\bibnamefont {Kato}},\ }\bibfield  {title}
  {\bibinfo {title} {{Systematic study on thermal conductivity of organic
  triangular lattice systems $\beta'$-X[Pd(dmit)$_2$]$_2$}},\ }\href
  {https://doi.org/10.1103/PhysRevB.105.245133} {\bibfield  {journal} {\bibinfo
   {journal} {Phys. Rev. B}\ }\textbf {\bibinfo {volume} {105}},\ \bibinfo
  {pages} {245133} (\bibinfo {year} {2022})}\BibitemShut {NoStop}%
\bibitem [{\citenamefont {Yamashita}\ \emph {et~al.}(2022)\citenamefont
  {Yamashita}, \citenamefont {Sato}, \citenamefont {Kasahara}, \citenamefont
  {Kasahara}, \citenamefont {Shibauchi},\ and\ \citenamefont
  {Matsuda}}]{Yamashita2022}%
  \BibitemOpen
  \bibfield  {author} {\bibinfo {author} {\bibfnamefont {M.}~\bibnamefont
  {Yamashita}}, \bibinfo {author} {\bibfnamefont {Y.}~\bibnamefont {Sato}},
  \bibinfo {author} {\bibfnamefont {Y.}~\bibnamefont {Kasahara}}, \bibinfo
  {author} {\bibfnamefont {S.}~\bibnamefont {Kasahara}}, \bibinfo {author}
  {\bibfnamefont {T.}~\bibnamefont {Shibauchi}},\ and\ \bibinfo {author}
  {\bibfnamefont {Y.}~\bibnamefont {Matsuda}},\ }\bibfield  {title} {\bibinfo
  {title} {{Resistivity and thermal conductivity of an organic insulator
  $\beta'$--EtMe$_3$Sb[Pd(dmit)$_2$]$_2$}},\ }\href
  {https://doi.org/10.1038/s41598-022-13155-8} {\bibfield  {journal} {\bibinfo
  {journal} {Sci. Rep.}\ }\textbf {\bibinfo {volume} {12}},\ \bibinfo {pages}
  {9187} (\bibinfo {year} {2022})}\BibitemShut {NoStop}%
\bibitem [{\citenamefont {Lewin}\ \emph {et~al.}(2023)\citenamefont {Lewin},
  \citenamefont {Frank}, \citenamefont {Ran}, \citenamefont {Paglione},\ and\
  \citenamefont {Butch}}]{Lewin_2023}%
  \BibitemOpen
  \bibfield  {author} {\bibinfo {author} {\bibfnamefont {S.~K.}\ \bibnamefont
  {Lewin}}, \bibinfo {author} {\bibfnamefont {C.~E.}\ \bibnamefont {Frank}},
  \bibinfo {author} {\bibfnamefont {S.}~\bibnamefont {Ran}}, \bibinfo {author}
  {\bibfnamefont {J.}~\bibnamefont {Paglione}},\ and\ \bibinfo {author}
  {\bibfnamefont {N.~P.}\ \bibnamefont {Butch}},\ }\bibfield  {title} {\bibinfo
  {title} {{A review of UTe$_2$ at high magnetic fields}},\ }\href
  {https://doi.org/10.1088/1361-6633/acfb93} {\bibfield  {journal} {\bibinfo
  {journal} {Rep. Prog. Phys.}\ }\textbf {\bibinfo {volume} {86}},\ \bibinfo
  {pages} {114501} (\bibinfo {year} {2023})}\BibitemShut {NoStop}%
\bibitem [{\citenamefont {Suetsugu}\ \emph {et~al.}(2024)\citenamefont
  {Suetsugu}, \citenamefont {Shimomura}, \citenamefont {Kamimura},
  \citenamefont {Asaba}, \citenamefont {Asaeda}, \citenamefont {Kosuge},
  \citenamefont {Sekino}, \citenamefont {Ikemori}, \citenamefont {Kasahara},
  \citenamefont {Kohsaka}, \citenamefont {Lee}, \citenamefont {Yanase},
  \citenamefont {Sakai}, \citenamefont {Opletal}, \citenamefont {Tokiwa},
  \citenamefont {Haga},\ and\ \citenamefont {Matsuda}}]{Suetsugu2024}%
  \BibitemOpen
  \bibfield  {author} {\bibinfo {author} {\bibfnamefont {S.}~\bibnamefont
  {Suetsugu}}, \bibinfo {author} {\bibfnamefont {M.}~\bibnamefont {Shimomura}},
  \bibinfo {author} {\bibfnamefont {M.}~\bibnamefont {Kamimura}}, \bibinfo
  {author} {\bibfnamefont {T.}~\bibnamefont {Asaba}}, \bibinfo {author}
  {\bibfnamefont {H.}~\bibnamefont {Asaeda}}, \bibinfo {author} {\bibfnamefont
  {Y.}~\bibnamefont {Kosuge}}, \bibinfo {author} {\bibfnamefont
  {Y.}~\bibnamefont {Sekino}}, \bibinfo {author} {\bibfnamefont
  {S.}~\bibnamefont {Ikemori}}, \bibinfo {author} {\bibfnamefont
  {Y.}~\bibnamefont {Kasahara}}, \bibinfo {author} {\bibfnamefont
  {Y.}~\bibnamefont {Kohsaka}}, \bibinfo {author} {\bibfnamefont
  {M.}~\bibnamefont {Lee}}, \bibinfo {author} {\bibfnamefont {Y.}~\bibnamefont
  {Yanase}}, \bibinfo {author} {\bibfnamefont {H.}~\bibnamefont {Sakai}},
  \bibinfo {author} {\bibfnamefont {P.}~\bibnamefont {Opletal}}, \bibinfo
  {author} {\bibfnamefont {Y.}~\bibnamefont {Tokiwa}}, \bibinfo {author}
  {\bibfnamefont {Y.}~\bibnamefont {Haga}},\ and\ \bibinfo {author}
  {\bibfnamefont {Y.}~\bibnamefont {Matsuda}},\ }\bibfield  {title} {\bibinfo
  {title} {{Fully gapped pairing state in spin-triplet superconductor
  UTe$_2$}},\ }\href {https://doi.org/10.1126/sciadv.adk3772} {\bibfield
  {journal} {\bibinfo  {journal} {Sci. Adv.}\ }\textbf {\bibinfo {volume}
  {10}},\ \bibinfo {pages} {eadk3772} (\bibinfo {year} {2024})}\BibitemShut
  {NoStop}%
\bibitem [{\citenamefont {Stankiewicz}\ \emph {et~al.}(2019)\citenamefont
  {Stankiewicz}, \citenamefont {Evangelisti}, \citenamefont {Rosa},
  \citenamefont {Schlottmann},\ and\ \citenamefont
  {Fisk}}]{PhysRevB.99.045138}%
  \BibitemOpen
  \bibfield  {author} {\bibinfo {author} {\bibfnamefont {J.}~\bibnamefont
  {Stankiewicz}}, \bibinfo {author} {\bibfnamefont {M.}~\bibnamefont
  {Evangelisti}}, \bibinfo {author} {\bibfnamefont {P.~F.~S.}\ \bibnamefont
  {Rosa}}, \bibinfo {author} {\bibfnamefont {P.}~\bibnamefont {Schlottmann}},\
  and\ \bibinfo {author} {\bibfnamefont {Z.}~\bibnamefont {Fisk}},\ }\bibfield
  {title} {\bibinfo {title} {{Physical properties of Sm$_x$B$_6$ single
  crystals}},\ }\href {https://doi.org/10.1103/PhysRevB.99.045138} {\bibfield
  {journal} {\bibinfo  {journal} {Phys. Rev. B}\ }\textbf {\bibinfo {volume}
  {99}},\ \bibinfo {pages} {045138} (\bibinfo {year} {2019})}\BibitemShut
  {NoStop}%
\bibitem [{\citenamefont {Sato}\ \emph {et~al.}(2019)\citenamefont {Sato},
  \citenamefont {Xiang}, \citenamefont {Kasahara}, \citenamefont {Taniguchi},
  \citenamefont {Kasahara}, \citenamefont {Chen}, \citenamefont {Asaba},
  \citenamefont {Tinsman}, \citenamefont {Murayama}, \citenamefont {Tanaka},
  \citenamefont {Mizukami}, \citenamefont {Shibauchi}, \citenamefont {Iga},
  \citenamefont {Singleton}, \citenamefont {Li},\ and\ \citenamefont
  {Matsuda}}]{Sato2019}%
  \BibitemOpen
  \bibfield  {author} {\bibinfo {author} {\bibfnamefont {Y.}~\bibnamefont
  {Sato}}, \bibinfo {author} {\bibfnamefont {Z.}~\bibnamefont {Xiang}},
  \bibinfo {author} {\bibfnamefont {Y.}~\bibnamefont {Kasahara}}, \bibinfo
  {author} {\bibfnamefont {T.}~\bibnamefont {Taniguchi}}, \bibinfo {author}
  {\bibfnamefont {S.}~\bibnamefont {Kasahara}}, \bibinfo {author}
  {\bibfnamefont {L.}~\bibnamefont {Chen}}, \bibinfo {author} {\bibfnamefont
  {T.}~\bibnamefont {Asaba}}, \bibinfo {author} {\bibfnamefont
  {C.}~\bibnamefont {Tinsman}}, \bibinfo {author} {\bibfnamefont
  {H.}~\bibnamefont {Murayama}}, \bibinfo {author} {\bibfnamefont
  {O.}~\bibnamefont {Tanaka}}, \bibinfo {author} {\bibfnamefont
  {Y.}~\bibnamefont {Mizukami}}, \bibinfo {author} {\bibfnamefont
  {T.}~\bibnamefont {Shibauchi}}, \bibinfo {author} {\bibfnamefont
  {F.}~\bibnamefont {Iga}}, \bibinfo {author} {\bibfnamefont {J.}~\bibnamefont
  {Singleton}}, \bibinfo {author} {\bibfnamefont {L.}~\bibnamefont {Li}},\ and\
  \bibinfo {author} {\bibfnamefont {Y.}~\bibnamefont {Matsuda}},\ }\bibfield
  {title} {\bibinfo {title} {{Unconventional thermal metallic state of
  charge-neutral fermions in an insulator}},\ }\href
  {https://doi.org/10.1038/s41567-019-0552-2} {\bibfield  {journal} {\bibinfo
  {journal} {Nat. Phys.}\ }\textbf {\bibinfo {volume} {15}},\ \bibinfo {pages}
  {954} (\bibinfo {year} {2019})}\BibitemShut {NoStop}%
\bibitem [{\citenamefont {Yamashita}\ \emph {et~al.}(2011)\citenamefont
  {Yamashita}, \citenamefont {Yamamoto}, \citenamefont {Nakazawa},
  \citenamefont {Tamura},\ and\ \citenamefont {Kato}}]{Yamashita2011}%
  \BibitemOpen
  \bibfield  {author} {\bibinfo {author} {\bibfnamefont {S.}~\bibnamefont
  {Yamashita}}, \bibinfo {author} {\bibfnamefont {T.}~\bibnamefont {Yamamoto}},
  \bibinfo {author} {\bibfnamefont {Y.}~\bibnamefont {Nakazawa}}, \bibinfo
  {author} {\bibfnamefont {M.}~\bibnamefont {Tamura}},\ and\ \bibinfo {author}
  {\bibfnamefont {R.}~\bibnamefont {Kato}},\ }\bibfield  {title} {\bibinfo
  {title} {{Gapless spin liquid of an organic triangular compound evidenced by
  thermodynamic measurements}},\ }\href {https://doi.org/10.1038/ncomms1274}
  {\bibfield  {journal} {\bibinfo  {journal} {Nat. Commun.}\ }\textbf {\bibinfo
  {volume} {2}},\ \bibinfo {pages} {275} (\bibinfo {year} {2011})}\BibitemShut
  {NoStop}%
\bibitem [{\citenamefont {Bourgeois-Hope}\ \emph {et~al.}(2019)\citenamefont
  {Bourgeois-Hope}, \citenamefont {Lalibert\'e}, \citenamefont
  {Lefran\ifmmode~\mbox{\c{c}}\else \c{c}\fi{}ois}, \citenamefont
  {Grissonnanche}, \citenamefont {de~Cotret}, \citenamefont {Gordon},
  \citenamefont {Kitou}, \citenamefont {Sawa}, \citenamefont {Cui},
  \citenamefont {Kato}, \citenamefont {Taillefer},\ and\ \citenamefont
  {Doiron-Leyraud}}]{PhysRevX.9.041051}%
  \BibitemOpen
  \bibfield  {author} {\bibinfo {author} {\bibfnamefont {P.}~\bibnamefont
  {Bourgeois-Hope}}, \bibinfo {author} {\bibfnamefont {F.}~\bibnamefont
  {Lalibert\'e}}, \bibinfo {author} {\bibfnamefont {E.}~\bibnamefont
  {Lefran\ifmmode~\mbox{\c{c}}\else \c{c}\fi{}ois}}, \bibinfo {author}
  {\bibfnamefont {G.}~\bibnamefont {Grissonnanche}}, \bibinfo {author}
  {\bibfnamefont {S.~R.}\ \bibnamefont {de~Cotret}}, \bibinfo {author}
  {\bibfnamefont {R.}~\bibnamefont {Gordon}}, \bibinfo {author} {\bibfnamefont
  {S.}~\bibnamefont {Kitou}}, \bibinfo {author} {\bibfnamefont
  {H.}~\bibnamefont {Sawa}}, \bibinfo {author} {\bibfnamefont {H.}~\bibnamefont
  {Cui}}, \bibinfo {author} {\bibfnamefont {R.}~\bibnamefont {Kato}}, \bibinfo
  {author} {\bibfnamefont {L.}~\bibnamefont {Taillefer}},\ and\ \bibinfo
  {author} {\bibfnamefont {N.}~\bibnamefont {Doiron-Leyraud}},\ }\bibfield
  {title} {\bibinfo {title} {{Thermal Conductivity of the Quantum Spin Liquid
  Candidate EtMe$_3$Sb[Pd(dmit)$_2$]$_2$: No Evidence of Mobile Gapless
  Excitations}},\ }\href {https://doi.org/10.1103/PhysRevX.9.041051} {\bibfield
   {journal} {\bibinfo  {journal} {Phys. Rev. X}\ }\textbf {\bibinfo {volume}
  {9}},\ \bibinfo {pages} {041051} (\bibinfo {year} {2019})}\BibitemShut
  {NoStop}%
\bibitem [{\citenamefont {Li}\ \emph {et~al.}(2024)\citenamefont {Li},
  \citenamefont {Xie}, \citenamefont {Huang}, \citenamefont {Zhuo},
  \citenamefont {Zhang}, \citenamefont {Choi}, \citenamefont {Wang},
  \citenamefont {Liang}, \citenamefont {Sun}, \citenamefont {Wu}, \citenamefont
  {Li}, \citenamefont {Zhou}, \citenamefont {Chen}, \citenamefont {Zhao},
  \citenamefont {Zhang},\ and\ \citenamefont {Sun}}]{PhysRevB.110.224414}%
  \BibitemOpen
  \bibfield  {author} {\bibinfo {author} {\bibfnamefont {N.}~\bibnamefont
  {Li}}, \bibinfo {author} {\bibfnamefont {M.~T.}\ \bibnamefont {Xie}},
  \bibinfo {author} {\bibfnamefont {Q.}~\bibnamefont {Huang}}, \bibinfo
  {author} {\bibfnamefont {Z.~W.}\ \bibnamefont {Zhuo}}, \bibinfo {author}
  {\bibfnamefont {Z.}~\bibnamefont {Zhang}}, \bibinfo {author} {\bibfnamefont
  {E.~S.}\ \bibnamefont {Choi}}, \bibinfo {author} {\bibfnamefont {Y.~Y.}\
  \bibnamefont {Wang}}, \bibinfo {author} {\bibfnamefont {H.}~\bibnamefont
  {Liang}}, \bibinfo {author} {\bibfnamefont {Y.}~\bibnamefont {Sun}}, \bibinfo
  {author} {\bibfnamefont {D.~D.}\ \bibnamefont {Wu}}, \bibinfo {author}
  {\bibfnamefont {Q.~J.}\ \bibnamefont {Li}}, \bibinfo {author} {\bibfnamefont
  {H.~D.}\ \bibnamefont {Zhou}}, \bibinfo {author} {\bibfnamefont
  {G.}~\bibnamefont {Chen}}, \bibinfo {author} {\bibfnamefont {X.}~\bibnamefont
  {Zhao}}, \bibinfo {author} {\bibfnamefont {Q.~M.}\ \bibnamefont {Zhang}},\
  and\ \bibinfo {author} {\bibfnamefont {X.~F.}\ \bibnamefont {Sun}},\
  }\bibfield  {title} {\bibinfo {title} {{Thermodynamics and heat transport in
  the quantum spin liquid candidates NaYbS$_2$ and NaYbSe$_2$}},\ }\href
  {https://doi.org/10.1103/PhysRevB.110.224414} {\bibfield  {journal} {\bibinfo
   {journal} {Phys. Rev. B}\ }\textbf {\bibinfo {volume} {110}},\ \bibinfo
  {pages} {224414} (\bibinfo {year} {2024})}\BibitemShut {NoStop}%
\bibitem [{\citenamefont {Ranjith}\ \emph {et~al.}(2019)\citenamefont
  {Ranjith}, \citenamefont {Luther}, \citenamefont {Reimann}, \citenamefont
  {Schmidt}, \citenamefont {Schlender}, \citenamefont {Sichelschmidt},
  \citenamefont {Yasuoka}, \citenamefont {Strydom}, \citenamefont {Skourski},
  \citenamefont {Wosnitza}, \citenamefont {K\"uhne}, \citenamefont {Doert},\
  and\ \citenamefont {Baenitz}}]{PhysRevB.100.224417}%
  \BibitemOpen
  \bibfield  {author} {\bibinfo {author} {\bibfnamefont {K.~M.}\ \bibnamefont
  {Ranjith}}, \bibinfo {author} {\bibfnamefont {S.}~\bibnamefont {Luther}},
  \bibinfo {author} {\bibfnamefont {T.}~\bibnamefont {Reimann}}, \bibinfo
  {author} {\bibfnamefont {B.}~\bibnamefont {Schmidt}}, \bibinfo {author}
  {\bibfnamefont {P.}~\bibnamefont {Schlender}}, \bibinfo {author}
  {\bibfnamefont {J.}~\bibnamefont {Sichelschmidt}}, \bibinfo {author}
  {\bibfnamefont {H.}~\bibnamefont {Yasuoka}}, \bibinfo {author} {\bibfnamefont
  {A.~M.}\ \bibnamefont {Strydom}}, \bibinfo {author} {\bibfnamefont
  {Y.}~\bibnamefont {Skourski}}, \bibinfo {author} {\bibfnamefont
  {J.}~\bibnamefont {Wosnitza}}, \bibinfo {author} {\bibfnamefont
  {H.}~\bibnamefont {K\"uhne}}, \bibinfo {author} {\bibfnamefont
  {T.}~\bibnamefont {Doert}},\ and\ \bibinfo {author} {\bibfnamefont
  {M.}~\bibnamefont {Baenitz}},\ }\bibfield  {title} {\bibinfo {title}
  {{Anisotropic field-induced ordering in the triangular-lattice quantum spin
  liquid NaYbSe$_2$}},\ }\href {https://doi.org/10.1103/PhysRevB.100.224417}
  {\bibfield  {journal} {\bibinfo  {journal} {Phys. Rev. B}\ }\textbf {\bibinfo
  {volume} {100}},\ \bibinfo {pages} {224417} (\bibinfo {year}
  {2019})}\BibitemShut {NoStop}%
\bibitem [{\citenamefont {Dai}\ \emph {et~al.}(2021)\citenamefont {Dai},
  \citenamefont {Zhang}, \citenamefont {Xie}, \citenamefont {Duan},
  \citenamefont {Gao}, \citenamefont {Zhu}, \citenamefont {Feng}, \citenamefont
  {Tao}, \citenamefont {Huang}, \citenamefont {Cao}, \citenamefont
  {Podlesnyak}, \citenamefont {Granroth}, \citenamefont {Everett},
  \citenamefont {Neuefeind}, \citenamefont {Voneshen}, \citenamefont {Wang},
  \citenamefont {Tan}, \citenamefont {Morosan}, \citenamefont {Wang},
  \citenamefont {Lin}, \citenamefont {Shu}, \citenamefont {Chen}, \citenamefont
  {Guo}, \citenamefont {Lu},\ and\ \citenamefont {Dai}}]{PhysRevX.11.021044}%
  \BibitemOpen
  \bibfield  {author} {\bibinfo {author} {\bibfnamefont {P.-L.}\ \bibnamefont
  {Dai}}, \bibinfo {author} {\bibfnamefont {G.}~\bibnamefont {Zhang}}, \bibinfo
  {author} {\bibfnamefont {Y.}~\bibnamefont {Xie}}, \bibinfo {author}
  {\bibfnamefont {C.}~\bibnamefont {Duan}}, \bibinfo {author} {\bibfnamefont
  {Y.}~\bibnamefont {Gao}}, \bibinfo {author} {\bibfnamefont {Z.}~\bibnamefont
  {Zhu}}, \bibinfo {author} {\bibfnamefont {E.}~\bibnamefont {Feng}}, \bibinfo
  {author} {\bibfnamefont {Z.}~\bibnamefont {Tao}}, \bibinfo {author}
  {\bibfnamefont {C.-L.}\ \bibnamefont {Huang}}, \bibinfo {author}
  {\bibfnamefont {H.}~\bibnamefont {Cao}}, \bibinfo {author} {\bibfnamefont
  {A.}~\bibnamefont {Podlesnyak}}, \bibinfo {author} {\bibfnamefont {G.~E.}\
  \bibnamefont {Granroth}}, \bibinfo {author} {\bibfnamefont {M.~S.}\
  \bibnamefont {Everett}}, \bibinfo {author} {\bibfnamefont {J.~C.}\
  \bibnamefont {Neuefeind}}, \bibinfo {author} {\bibfnamefont {D.}~\bibnamefont
  {Voneshen}}, \bibinfo {author} {\bibfnamefont {S.}~\bibnamefont {Wang}},
  \bibinfo {author} {\bibfnamefont {G.}~\bibnamefont {Tan}}, \bibinfo {author}
  {\bibfnamefont {E.}~\bibnamefont {Morosan}}, \bibinfo {author} {\bibfnamefont
  {X.}~\bibnamefont {Wang}}, \bibinfo {author} {\bibfnamefont {H.-Q.}\
  \bibnamefont {Lin}}, \bibinfo {author} {\bibfnamefont {L.}~\bibnamefont
  {Shu}}, \bibinfo {author} {\bibfnamefont {G.}~\bibnamefont {Chen}}, \bibinfo
  {author} {\bibfnamefont {Y.}~\bibnamefont {Guo}}, \bibinfo {author}
  {\bibfnamefont {X.}~\bibnamefont {Lu}},\ and\ \bibinfo {author}
  {\bibfnamefont {P.}~\bibnamefont {Dai}},\ }\bibfield  {title} {\bibinfo
  {title} {{Spinon Fermi Surface Spin Liquid in a Triangular Lattice
  Antiferromagnet NaYbSe$_2$}},\ }\href
  {https://doi.org/10.1103/PhysRevX.11.021044} {\bibfield  {journal} {\bibinfo
  {journal} {Phys. Rev. X}\ }\textbf {\bibinfo {volume} {11}},\ \bibinfo
  {pages} {021044} (\bibinfo {year} {2021})}\BibitemShut {NoStop}%
\bibitem [{\citenamefont {Yu}\ \emph {et~al.}(2017)\citenamefont {Yu},
  \citenamefont {Xu}, \citenamefont {He}, \citenamefont {Kratochvilova},
  \citenamefont {Huang}, \citenamefont {Ni}, \citenamefont {Wang},
  \citenamefont {Cheong}, \citenamefont {Park},\ and\ \citenamefont
  {Li}}]{PhysRevB.96.081111}%
  \BibitemOpen
  \bibfield  {author} {\bibinfo {author} {\bibfnamefont {Y.~J.}\ \bibnamefont
  {Yu}}, \bibinfo {author} {\bibfnamefont {Y.}~\bibnamefont {Xu}}, \bibinfo
  {author} {\bibfnamefont {L.~P.}\ \bibnamefont {He}}, \bibinfo {author}
  {\bibfnamefont {M.}~\bibnamefont {Kratochvilova}}, \bibinfo {author}
  {\bibfnamefont {Y.~Y.}\ \bibnamefont {Huang}}, \bibinfo {author}
  {\bibfnamefont {J.~M.}\ \bibnamefont {Ni}}, \bibinfo {author} {\bibfnamefont
  {L.}~\bibnamefont {Wang}}, \bibinfo {author} {\bibfnamefont {S.-W.}\
  \bibnamefont {Cheong}}, \bibinfo {author} {\bibfnamefont {J.-G.}\
  \bibnamefont {Park}},\ and\ \bibinfo {author} {\bibfnamefont {S.~Y.}\
  \bibnamefont {Li}},\ }\bibfield  {title} {\bibinfo {title} {{Heat transport
  study of the spin liquid candidate 1T-TaS$_2$}},\ }\href
  {https://doi.org/10.1103/PhysRevB.96.081111} {\bibfield  {journal} {\bibinfo
  {journal} {Phys. Rev. B}\ }\textbf {\bibinfo {volume} {96}},\ \bibinfo
  {pages} {081111} (\bibinfo {year} {2017})}\BibitemShut {NoStop}%
\bibitem [{\citenamefont {Kratochvilova}\ \emph {et~al.}(2017)\citenamefont
  {Kratochvilova}, \citenamefont {Hillier}, \citenamefont {Wildes},
  \citenamefont {Wang}, \citenamefont {Cheong},\ and\ \citenamefont
  {Park}}]{Kratochvilova2017}%
  \BibitemOpen
  \bibfield  {author} {\bibinfo {author} {\bibfnamefont {M.}~\bibnamefont
  {Kratochvilova}}, \bibinfo {author} {\bibfnamefont {A.~D.}\ \bibnamefont
  {Hillier}}, \bibinfo {author} {\bibfnamefont {A.~R.}\ \bibnamefont {Wildes}},
  \bibinfo {author} {\bibfnamefont {L.}~\bibnamefont {Wang}}, \bibinfo {author}
  {\bibfnamefont {S.-W.}\ \bibnamefont {Cheong}},\ and\ \bibinfo {author}
  {\bibfnamefont {J.-G.}\ \bibnamefont {Park}},\ }\bibfield  {title} {\bibinfo
  {title} {{The low-temperature highly correlated quantum phase in the
  charge-density-wave 1T-TaS$_2$ compound}},\ }\href
  {https://doi.org/10.1038/s41535-017-0048-1} {\bibfield  {journal} {\bibinfo
  {journal} {npj Quantum Mater.}\ }\textbf {\bibinfo {volume} {2}},\ \bibinfo
  {pages} {42} (\bibinfo {year} {2017})}\BibitemShut {NoStop}%
\bibitem [{\citenamefont {Imajo}\ \emph {et~al.}(2019)\citenamefont {Imajo},
  \citenamefont {Kohama}, \citenamefont {Miyake}, \citenamefont {Dong},
  \citenamefont {Tokunaga}, \citenamefont {Flouquet}, \citenamefont {Kindo},\
  and\ \citenamefont {Aoki}}]{Imajo2019}%
  \BibitemOpen
  \bibfield  {author} {\bibinfo {author} {\bibfnamefont {S.}~\bibnamefont
  {Imajo}}, \bibinfo {author} {\bibfnamefont {Y.}~\bibnamefont {Kohama}},
  \bibinfo {author} {\bibfnamefont {A.}~\bibnamefont {Miyake}}, \bibinfo
  {author} {\bibfnamefont {C.}~\bibnamefont {Dong}}, \bibinfo {author}
  {\bibfnamefont {M.}~\bibnamefont {Tokunaga}}, \bibinfo {author}
  {\bibfnamefont {J.}~\bibnamefont {Flouquet}}, \bibinfo {author}
  {\bibfnamefont {K.}~\bibnamefont {Kindo}},\ and\ \bibinfo {author}
  {\bibfnamefont {D.}~\bibnamefont {Aoki}},\ }\bibfield  {title} {\bibinfo
  {title} {{Thermodynamic Investigation of Metamagnetism in Pulsed High
  Magnetic Fields on Heavy Fermion Superconductor UTe$_2$}},\ }\href
  {https://doi.org/10.7566/JPSJ.88.083705} {\bibfield  {journal} {\bibinfo
  {journal} {J. Phys. Soc. Jpn.}\ }\textbf {\bibinfo {volume} {88}},\ \bibinfo
  {pages} {083705} (\bibinfo {year} {2019})}\BibitemShut {NoStop}%
\bibitem [{\citenamefont {Ishihara}\ \emph {et~al.}(2023)\citenamefont
  {Ishihara}, \citenamefont {Roppongi}, \citenamefont {Kobayashi},
  \citenamefont {Imamura}, \citenamefont {Mizukami}, \citenamefont {Sakai},
  \citenamefont {Opletal}, \citenamefont {Tokiwa}, \citenamefont {Haga},
  \citenamefont {Hashimoto},\ and\ \citenamefont {Shibauchi}}]{Ishihara2023}%
  \BibitemOpen
  \bibfield  {author} {\bibinfo {author} {\bibfnamefont {K.}~\bibnamefont
  {Ishihara}}, \bibinfo {author} {\bibfnamefont {M.}~\bibnamefont {Roppongi}},
  \bibinfo {author} {\bibfnamefont {M.}~\bibnamefont {Kobayashi}}, \bibinfo
  {author} {\bibfnamefont {K.}~\bibnamefont {Imamura}}, \bibinfo {author}
  {\bibfnamefont {Y.}~\bibnamefont {Mizukami}}, \bibinfo {author}
  {\bibfnamefont {H.}~\bibnamefont {Sakai}}, \bibinfo {author} {\bibfnamefont
  {P.}~\bibnamefont {Opletal}}, \bibinfo {author} {\bibfnamefont
  {Y.}~\bibnamefont {Tokiwa}}, \bibinfo {author} {\bibfnamefont
  {Y.}~\bibnamefont {Haga}}, \bibinfo {author} {\bibfnamefont {K.}~\bibnamefont
  {Hashimoto}},\ and\ \bibinfo {author} {\bibfnamefont {T.}~\bibnamefont
  {Shibauchi}},\ }\bibfield  {title} {\bibinfo {title} {{Chiral
  superconductivity in UTe$_2$ probed by anisotropic low-energy excitations}},\
  }\href {https://doi.org/10.1038/s41467-023-38688-y} {\bibfield  {journal}
  {\bibinfo  {journal} {Nat. Commun.}\ }\textbf {\bibinfo {volume} {14}},\
  \bibinfo {pages} {2966} (\bibinfo {year} {2023})}\BibitemShut {NoStop}%
\bibitem [{\citenamefont {Bordelon}\ \emph {et~al.}(2019)\citenamefont
  {Bordelon}, \citenamefont {Kenney}, \citenamefont {Liu}, \citenamefont
  {Hogan}, \citenamefont {Posthuma}, \citenamefont {Kavand}, \citenamefont
  {Lyu}, \citenamefont {Sherwin}, \citenamefont {Butch}, \citenamefont {Brown},
  \citenamefont {Graf}, \citenamefont {Balents},\ and\ \citenamefont
  {Wilson}}]{Bordelon2019}%
  \BibitemOpen
  \bibfield  {author} {\bibinfo {author} {\bibfnamefont {M.~M.}\ \bibnamefont
  {Bordelon}}, \bibinfo {author} {\bibfnamefont {E.}~\bibnamefont {Kenney}},
  \bibinfo {author} {\bibfnamefont {C.}~\bibnamefont {Liu}}, \bibinfo {author}
  {\bibfnamefont {T.}~\bibnamefont {Hogan}}, \bibinfo {author} {\bibfnamefont
  {L.}~\bibnamefont {Posthuma}}, \bibinfo {author} {\bibfnamefont
  {M.}~\bibnamefont {Kavand}}, \bibinfo {author} {\bibfnamefont
  {Y.}~\bibnamefont {Lyu}}, \bibinfo {author} {\bibfnamefont {M.}~\bibnamefont
  {Sherwin}}, \bibinfo {author} {\bibfnamefont {N.~P.}\ \bibnamefont {Butch}},
  \bibinfo {author} {\bibfnamefont {C.}~\bibnamefont {Brown}}, \bibinfo
  {author} {\bibfnamefont {M.~J.}\ \bibnamefont {Graf}}, \bibinfo {author}
  {\bibfnamefont {L.}~\bibnamefont {Balents}},\ and\ \bibinfo {author}
  {\bibfnamefont {S.~D.}\ \bibnamefont {Wilson}},\ }\bibfield  {title}
  {\bibinfo {title} {{Field-tunable quantum disordered ground state in the
  triangular-lattice antiferromagnet NaYbO$_2$}},\ }\href
  {https://doi.org/10.1038/s41567-019-0594-5} {\bibfield  {journal} {\bibinfo
  {journal} {Nat. Phys.}\ }\textbf {\bibinfo {volume} {15}},\ \bibinfo {pages}
  {1058} (\bibinfo {year} {2019})}\BibitemShut {NoStop}%
\bibitem [{\citenamefont {Vollmer}(2009)}]{Vollmer_2009}%
  \BibitemOpen
  \bibfield  {author} {\bibinfo {author} {\bibfnamefont {M.}~\bibnamefont
  {Vollmer}},\ }\bibfield  {title} {\bibinfo {title} {Newton's law of cooling
  revisited},\ }\href {https://doi.org/10.1088/0143-0807/30/5/014} {\bibfield
  {journal} {\bibinfo  {journal} {Eur. J. Phys.}\ }\textbf {\bibinfo {volume}
  {30}},\ \bibinfo {pages} {1063} (\bibinfo {year} {2009})}\BibitemShut
  {NoStop}%
\bibitem [{\citenamefont {Kaka{\c{c}}}\ and\ \citenamefont
  {Yener}(1985)}]{Kakac1985}%
  \BibitemOpen
  \bibfield  {author} {\bibinfo {author} {\bibfnamefont {S.}~\bibnamefont
  {Kaka{\c{c}}}}\ and\ \bibinfo {author} {\bibfnamefont {Y.}~\bibnamefont
  {Yener}},\ }\href {https://books.google.com/books?id=tQJRAAAAMAAJ} {\emph
  {\bibinfo {title} {Heat Conduction}}}\ (\bibinfo  {publisher} {Hemisphere
  Publishing Corporation},\ \bibinfo {year} {1985})\BibitemShut {NoStop}%
\bibitem [{\citenamefont {Mills}(1992)}]{Mills1992}%
  \BibitemOpen
  \bibfield  {author} {\bibinfo {author} {\bibfnamefont {A.}~\bibnamefont
  {Mills}},\ }\href {https://books.google.com/books?id=IVzSHjZ2LeEC} {\emph
  {\bibinfo {title} {Heat Transfer}}}\ (\bibinfo  {publisher} {Irwin},\
  \bibinfo {year} {1992})\BibitemShut {NoStop}%
\bibitem [{\citenamefont {Chen}\ \emph {et~al.}(2015)\citenamefont {Chen},
  \citenamefont {Fu},\ and\ \citenamefont {Xu}}]{Chen2015}%
  \BibitemOpen
  \bibfield  {author} {\bibinfo {author} {\bibfnamefont {Q.}~\bibnamefont
  {Chen}}, \bibinfo {author} {\bibfnamefont {R.-H.}\ \bibnamefont {Fu}},\ and\
  \bibinfo {author} {\bibfnamefont {Y.-C.}\ \bibnamefont {Xu}},\ }\bibfield
  {title} {\bibinfo {title} {Electrical circuit analogy for heat transfer
  analysis and optimization in heat exchanger networks},\ }\href
  {https://doi.org/https://doi.org/10.1016/j.apenergy.2014.11.021} {\bibfield
  {journal} {\bibinfo  {journal} {Appl. Energy}\ }\textbf {\bibinfo {volume}
  {139}},\ \bibinfo {pages} {81} (\bibinfo {year} {2015})}\BibitemShut
  {NoStop}%
\bibitem [{\citenamefont {Bueno}\ \emph {et~al.}(2012)\citenamefont {Bueno},
  \citenamefont {Norford}, \citenamefont {Pigeon},\ and\ \citenamefont
  {Britter}}]{Bueno2012}%
  \BibitemOpen
  \bibfield  {author} {\bibinfo {author} {\bibfnamefont {B.}~\bibnamefont
  {Bueno}}, \bibinfo {author} {\bibfnamefont {L.}~\bibnamefont {Norford}},
  \bibinfo {author} {\bibfnamefont {G.}~\bibnamefont {Pigeon}},\ and\ \bibinfo
  {author} {\bibfnamefont {R.}~\bibnamefont {Britter}},\ }\bibfield  {title}
  {\bibinfo {title} {A resistance-capacitance network model for the analysis of
  the interactions between the energy performance of buildings and the urban
  climate},\ }\href
  {https://doi.org/https://doi.org/10.1016/j.buildenv.2012.01.023} {\bibfield
  {journal} {\bibinfo  {journal} {Build. Environ.}\ }\textbf {\bibinfo {volume}
  {54}},\ \bibinfo {pages} {116} (\bibinfo {year} {2012})}\BibitemShut
  {NoStop}%
\bibitem [{\citenamefont {Silva}(2022)}]{Silva2022}%
  \BibitemOpen
  \bibfield  {author} {\bibinfo {author} {\bibfnamefont {D.}~\bibnamefont
  {Silva}},\ }\bibfield  {title} {\bibinfo {title} {{Modeling the Transient
  Response of Thermal Circuits}},\ }\bibfield  {journal} {\bibinfo  {journal}
  {Appl. Sci.}\ }\textbf {\bibinfo {volume} {12}},\ \href
  {https://doi.org/10.3390/app122412555} {10.3390/app122412555} (\bibinfo
  {year} {2022})\BibitemShut {NoStop}%
\bibitem [{sup()}]{supp}%
  \BibitemOpen
  \href@noop {} {\bibinfo  {journal} {See the Supplementary Material for more
  details}\ }\BibitemShut {NoStop}%
\end{thebibliography}%


\clearpage
\pagebreak
\onecolumngrid

\begin{center}
\textbf{\large Supplementary material \--- Resolving the Thermal Paradox: Many-body localization or fractionalization?}\\[.2cm]
Saikat Banerjee$^{1,\,*}$, Piers Coleman$^{1,2,\,\dag}$ \\[.1cm]
{\itshape 
${}^1$ Center for Materials Theory, Rutgers University, Piscataway, New Jersey, 08854, USA \\
${}^2$ Hubbard Theory Consortium, Department of Physics, \\
Royal Holloway, University of London, Egham, Surrey TW20 0EX, UK \\}
(Dated: \today)\\[1cm]
\end{center}

\setcounter{equation}{0}
\setcounter{figure}{0}
\setcounter{table}{0}
\setcounter{page}{1}
\renewcommand{\theequation}{S\arabic{equation}}
\renewcommand{\thefigure}{S\arabic{figure}}
\renewcommand{\bibnumfmt}[1]{[S#1]}
\renewcommand{\citenumfont}[1]{S#1}

\clearpage

\section{Impedance analysis \label{sec:sec_1}}

Here, we provide the derivation of time constants for the various thermal circuits relevant to the measurement of the specific heat and the thermal conductivity in the main text. The impedance for the first circuit in Fig.~1(a) in the main text is calculated as follows. First, we label the imaginary frequency by $-i\omega \rightarrow s$. It is clear that the impedance for $C_{\rm{f}}$, and the thermal resistance $R_{\rm{pf}}$ is in series, which together is in parallel to the capacitance $C_{\rm{p}}$. This altogether yields a total impedance $Z_1(s)$ as 
\begin{equation}\label{seq.1}
Z_1(s) 
= 
\frac{1}{sC_{\rm{p}} + \frac{1}{R_{\rm{pf}} + \frac{1}{sC_{\rm{f}}}}}
=
\frac{1}{sC_{\rm{p}}} \frac{\left(s + \frac{1}{R_{\rm{pf}} C_{\rm{f}}}\right)}{s + \frac{1}{R_{\rm{pf}} C_{\rm{p}}} + \frac{1}{R_{\rm{pf}} C_{\rm{f}}}},
\end{equation}
which gives a pole at 
\begin{equation}\label{seq.2}
-s_0 = \frac{1}{\tau_0} = \frac{1}{R_{\rm{pf}} C_{\rm{p}}} + \frac{1}{R_{\rm{pf}} C_{\rm{f}}}.
\end{equation}
This is consistent with the time constant relevant to the specific heat measurement, as mentioned in Eq.~\eqref{eq.4} in the main text. Note that the other time constant is infinite [$s = 0$ in Eq.~\eqref{seq.2}]. \\

\textbf{Impedance for thermal conductivity} \--- Next, we compute the impedance for the second circuit in the main text [see Fig.~1(b)] in the same manner. In this case, though, two distinct time constants emerge. The impedance for the thermal conductivity measurements is calculated as
\begin{equation}\label{seq.3}
Z_2(s) 
=
\frac{1}{sC_{\rm{p}}+ \frac{1}{Z_{R}(s)}+ \frac{1}{R_{\rm{p}}}}, \quad
\text{where}, 
\quad
Z_R(s)
=
R_{\rm{pf}}+ \frac{1}{s C_{\rm{f}} + \frac{1}{R_{\rm{f}}}}.
\end{equation}
We now write the form of $Z_2(s)$ explicitly to obtain the corresponding pole structure. The latter is obtained as
\begin{equation}\label{seq.4}
Z_{2}(s) 
= 
\frac{R_{\rm{p}} (C_{\rm{f}} R_{\rm{f}} R_{\rm{pf}} s + R_{\rm{pf}} + R_{\rm{f}})}
{C_{\rm{p}} C_{\rm{f}} R_{\rm{p}} R_{\rm{pf}} R_{\rm{f}} s^2 + s 
[C_{\rm{f}} (R_{\rm{p}} + R_{\rm{pf}} ) R_{\rm{f}} + C_{\rm{p}} R_{\rm{p}} (R_{\rm{pf}} + R_{\rm{f}})] + R_{\rm{p}} + R_{\rm{pf}} + R_{\rm{f}}},
\end{equation}
which subsequently provides the two decay rates by solving for the roots of the denominator in Eq.~\eqref{seq.4}. Simplifying further, we obtain the roots as
\begin{subequations}
\begin{align}
\label{seq.5.1}
s_1 
&
= 
\sqrt{
\left(
\frac{\frac{1}{R_{\rm{p}}} + \frac{1}{R_{\rm{pf}}}} {2 C_{\rm{p}}}
-
\frac{\frac{1}{R_{\rm{pf}}} + \frac{1}{R_{\rm{f}}}}{2C_{\rm{f}}}
\right)^2
+
\frac{1}{C_{\rm{p}} C_{\rm{f}} R_{\rm{pf}}^2}}
-
\frac{\frac{1}{R_{\rm{p}}} + \frac{1}{R_{\rm{pf}}}}{2C_{\rm{p}}} 
- 
\frac{\frac{1}{R_{\rm{pf}}} + \frac{1}{R_{\rm{f}}}}{2 C_{\rm{f}}}, \\
\label{seq.5.2}
s_{2}
&
=
-
\sqrt{
\left(
\frac{\frac{1}{R_{\rm{p}}} + \frac{1}{R_{\rm{pf}}}} {2 C_{\rm{p}}}
-
\frac{\frac{1}{R_{\rm{pf}}} + \frac{1}{R_{\rm{f}}}}{2C_{\rm{f}}}
\right)^2
+
\frac{1}{C_{\rm{p}} C_{\rm{f}} R_{\rm{pf}}^2}}
-
\frac{\frac{1}{R_{\rm{p}}} + \frac{1}{R_{\rm{pf}}}}{2C_{\rm{p}}} 
- 
\frac{\frac{1}{R_{\rm{pf}}} + \frac{1}{R_{\rm{f}}}}{2 C_{\rm{f}}}.
\end{align}
\end{subequations}
The denominator in Eq.~\eqref{seq.4} can be written as $s^{2} + \alpha s + \beta $, where we have 
\begin{equation}\label{seq.6}
\alpha
= 
\frac{1}{R_{\rm{pf}}}
\left(
\frac{R_{\rm{p}} + R_{\rm{pf}}}{R_{\rm{p}} C_{\rm{p}}} 
+ 
\frac{R_{\rm{pf}} + R_{\rm{pf}}}{R_{\rm{pf}} C_{\rm{f}}} 
\right),
\quad
\beta
=
\frac{R_{\rm{p}} + R_{\rm{pf}} + R_{\rm{f}}}{C_{\rm{p}} C_{\rm{f}} R_{\rm{p}} R_{\rm{pf}} R_{\rm{f}}}.
\end{equation} \\

\textbf{Slow and fast poles} \textbf{\---} Now we analyze the structure of the poles at various limiting cases. Typically, the $C_{\rm{f}} \gg C_{\rm{p}}$ in the various materials mentioned in the main text. Therefore, in this case, we have the $\beta \to 0$, and we have the fast pole at 
\begin{equation}\label{seq.7}
s_1 
=
- 
\frac{1}{\tau_{\rm{F}}} 
= 
-\alpha 
\sim
-\frac{R_{\rm{p}} + R_{\rm{pf}}} {R_{\rm{p}} R_{\rm{pf}} C_{\rm{p}}}
\quad
\Rightarrow
\quad
\tau_{\rm{F}}
\sim
C_{\rm{p}}
\left(
\frac{R_{\rm{p}} R_{\rm{pf}}} {R_{\rm{p}} + R_{\rm{pf}}} 
\right).
\end{equation}
On the other hand, since $s_1 s_2 = \beta$, we have a slow pole leading to slow time-constant as 
\begin{equation}\label{seq.8}
s_2 
= 
-\frac{1}{\tau_{\rm{S}}}
\sim
\frac{
R_{\rm{p}} + R_{\rm{pf}} + R_{\rm{f}}}
{C_{\rm{p}} C_{\rm{f}} (R_{\rm{p}} R_{\rm{pf}} R_{\rm{f}})}
\left(
\frac{
C_{\rm{p}} R_{\rm{p}} R_{\rm{pf}}}{-(R_{\rm{p}} + R_{\rm{pf}})}
\right) 
= 
-\frac{R_{\rm{pf}}+R_{\rm{p}}+R_{\rm{f}}}{C_{\rm{f}}R_{\rm{f}} (R_{\rm{pf}}+R_{\rm{p}})},
\quad
\Rightarrow
\quad
\tau_{\rm{S}}
= 
C_{\rm{f}} 
\left(
\frac{1}{\frac{1}{R_{\rm{f}}} + \frac{1}{R_{\rm{p}} + R_{\rm{pf}}}}
\right).
\end{equation}
These two time constants are inherent to the thermal conductivity measurement circuits as explained in Eq.~\eqref{eq.5} in the main text. We now move on to deriving the time-dependent transient temperature profile in the sample using Laplace transformation. \\

\textbf{Response in time domain} \textbf{\---} We can formaly write the impedance in Eq.~\eqref{seq.4} as following
\begin{equation}\label{seq.9}
Z_2(s) = \frac{s + s_0}{C_{\rm{ph}} (s-s_1) (s-s_2)},
\quad
\text{where}
\quad
s_0 = \frac{1}{C_{\rm{f}}} 
\left(
\frac{1}{R_{\rm{pf}}} + \frac{1}{R_{\rm{f}}} 
\right),
\end{equation}
and $s_1$, $s_2$ are the poles as defined in Eqs.~\eqref{seq.5.1},\eqref{seq.5.2}. Simplifying further we rewrite Eq.~\eqref{seq.9} as 
\begin{equation}\label{seq.10}
Z_2(s) 
=  
\frac{1}{C_{\rm{p}}} 
\left[ 
\frac{s_1 + s_0}{(s-s_1)(s_1-s_2)}
-
\frac{s_2 + s_0}{(s-s_2)(s_1-s_2) }
\right].
\end{equation}
Now, we utilize the electrical analogy of Laplace transformation to obtain the temperature profile as $T(s) = Z_2(s) I(s)$, where $I(s)$ is the thermal current flowing in the sample. We adopt a crude approximation to the thermal current as 
\begin{equation}\label{seq.11}
I(s) = I_0 \int_0^\infty e^{-st} dt = \frac{I_0}{s},
\end{equation}
where $I_0$ is the amplitude of the steady-state thermal current at a long time scale. Plugging this back into the temperature profile and simplifying it further we have
\begin{equation}\label{seq.12}
T(s) 
= 
I_0 
\frac{Z_2(s)}{s}
= 
\frac{I_0}{C_{\rm{p}}} 
\left[ 
	\frac{s_1 + s_0}{s_1(s_1-s_2)}
		\left(
			\frac{1}{s-s_1} - \frac{1}{s}
		\right)
	- 
	\frac{s_2 + s_0}{s_2(s_1-s_2)}
		\left(
			\frac{1}{s-s_2}-\frac{1}{s}
		\right)
\right].
\end{equation}
Utilizing the Laplace transformation of $e^{at}$ as $\mathcal{L}[e^{at}] = \tfrac{1}{s-a}$, we perform the inverse Laplace transformation on Eq.~\eqref{seq.12} to obtain the temperature profile as 
\begin{equation}\label{seq.13}
T(t) 
= 
\frac{I_0}{C_{\rm{p}}} 
\left[ 
	-\frac{s_1 + s_0}{s_1(s_1-s_2)}
		(1-e^{s_1 t})
		+ 
	\frac{s_2 + s_0}{s_2(s_1-s_2)}
		(1-e^{s_2t})
\right].
\end{equation}
where we labeled two constants $A_1$, and $A_2$ as defined in Eq.~\eqref{eq.6} of the main text. They are defined in terms of the poles as
\begin{equation}\label{seq.14}
A_1 = -\frac{s_1 + s_0}{C_{\rm{p}} s_1(s_1-s_2)}, 
\quad 
A_2 = \frac{s_1 + s_0}{C_{\rm{p}} s_2(s_1-s_2)}, 
\quad
\text{and} 
\quad
\tau_{1,2} = - \frac{1}{s_{1,2}}.
\end{equation}


\end{document}